\documentclass[onecolumn]{aastex631}

\newcommand{\bp}[1]{#1}

\begin{document}

\title{Beyond Point Masses. IV. TNO Altjira is Likely a Hierarchical Triple Discovered Through Non-Keplerian Motion}

\author[0000-0001-8780-8480]{Maia A Nelsen}
\affiliation{Brigham Young University, Department of Physics and Astronomy, N283 ESC, Provo, UT 84602, USA}

\author[0000-0003-1080-9770]{Darin Ragozzine}
\affiliation{Brigham Young University, Department of Physics and Astronomy, N283 ESC, Provo, UT 84602, USA}

\author[0000-0002-1788-870X]{Benjamin C. N. Proudfoot}
\affiliation{Brigham Young University, Department of Physics and Astronomy, N283 ESC, Provo, UT 84602, USA}
\affiliation{Florida Space Institute, University of Central Florida, 12354 Research Parkway, Orlando, FL 32826, USA}

\author[0000-0001-6838-1530]{William G. Giforos}
\affiliation{Brigham Young University, Department of Physics and Astronomy, N283 ESC, Provo, UT 84602, USA}

\author[0000-0002-8296-6540]{Will Grundy}
\affiliation{Lowell Observatory, Flagstaff, AZ 86001, USA}

\begin{abstract}

Dynamically studying Trans-Neptunian Object (TNO) binaries allows us to measure masses and orbits. Most of the known objects appear to have only two components, except (47171) Lempo which is the single known hierarchical triple system with three similar-mass components. Though hundreds of TNOs have been imaged with high-resolution telescopes, no other hierarchical triples (or trinaries) have been found among solar system small bodies, even though they are predicted in planetesimal formation models such as gravitational collapse after the streaming instability. By going beyond the point-mass assumption and modeling TNO orbits as non-Keplerian, we open a new window into the shapes and spins of the components, including the possible presence of unresolved ``inner'' binaries. Here we present evidence for a new hierarchical triple, (148780) Altjira (2001 UQ$_{18}$), based on non-Keplerian dynamical modeling of the two observed components. We incorporate two recent Hubble Space Telescope (HST) observations, leading to a 17 year observational baseline. We present a new open-source Bayesian Point Spread Function (PSF) fitting code called \texttt{nPSF} that provides precise relative astrometry and uncertainties for single images. Our non-Keplerian analysis measures a statistically-significant ($\sim$2.5-$\sigma$) non-spherical shape for Altjira. The measured $J_2$ is best explained as an unresolved inner binary and an example hierarchical triple model gives the best fit to the observed astrometry. Using an updated non-Keplerian ephemeris (which is significantly different from the Keplerian predictions), we show that the predicted mutual event season for Altjira has already begun with several excellent opportunities for observations through $\sim$2030.  
\end{abstract}


\section{Introduction and Background} \label{sec:intro}

Some of the best observational constraints on the streaming instability (SI) hypothesis of planetesimal formation come from the orbital properties of Trans-Neptunian Object (TNO) binaries. In particular, models of post-SI gravitational collapse are able to match many properties of TNO binary orbits like the orientation and the wide separations with near equal mass ratios \citep[e.g.][]{2019NatAs...3..808N}. These binaries are prevalent \citep{2019Icar..334...62G} in the ``cold classical'' portion of the Kuiper belt where heliocentric inclinations are very low \citep[especially when measured in dynamically meaningful terms of proper/free inclinations relative to the local forcing plane, see][]{2022ApJS..259...54H}. The cold classical population is also thought to have formed \emph{in situ} and NASA's New Horizons mission observations of cold classical Arrokoth points to a specific example of SI-formation \citep[e.g.,][]{2020Sci...367.6620M}. 

Dynamically, much of the post-SI gravitational collapse process is about the evolution and flow of angular momentum. The shrinking cloud soon has an excess of angular momentum (per unit mass) which naturally forms wide near-equal mass binaries. However, some of this angular momentum cascades down to the components of these binaries which provides the initial conditions for the spins and shapes of the components. Most gravitational collapse models use effective particle sizes and other approximations that make it challenging to resolve the angular momenta of the individual components \citep{2010AJ....140..785N,2020A&A...643A..55R,2021PSJ.....2...27N}, Recent work by \citet{2023ApJ...943..125P} shows that collapsing pebble clouds in the outer solar system frequently form into many components, many of which are near-equal binaries in close or contact configurations. Future generations of models will hopefully consider the evolution of these systems beyond initial formation to determine how angular momentum is partitioned into the reservoirs of wide binary and two spins as this is an important observational constraint for future models. 

Some models already show the components of the main binary can have so much angular momentum that one or both of the individual components should itself be a binary \citep{2010AJ....140..785N,2020A&A...643A..55R,2021PSJ.....2...27N}. We would then distinguish between the ``outer'' binary and an inner binary (or binaries). The ``hierarchical triple'' or ``hierarchical quadruple'' configuration is a relatively dynamically stable endpoint even for near-equal size components, as seen in stellar multiple systems. Presently, there is only one observational example of a hierarchical triple (sometimes called ``trinary'') configuration in the solar system: the TNO (47171) Lempo. 

Originally thought to be a binary, Lempo was barely resolved as a triple by the highest resolution Hubble Space Telescope (HST) camera by \citet{benecchi201047171}. A non-Keplerian three-point-mass model for the system shows an inner binary of the two larger ($\sim$250 km diameter) objects separated by $\sim$850 km orbited by a smaller component ($\sim$125 km) with a semi-major axis of $\sim$7600 km (Ragozzine et al., submitted, hereafter Paper I). This configuration has about as much angular momentum in the inner binary as in the outer binary. The three-point-mass fit also preserves the significant eccentricities ($\sim$0.12 for the inner binary, $\sim$0.29 for the outer binary) and orientation that \citet{correia2018chaotic} found to be dynamically unstable on timescales orders of magnitude shorter than the age of the solar system. Paper I suggests that further work is needed to characterize and explain the Lempo system and its long-term evolution. 

Despite being a relatively massive system, Lempo was barely resolved (0.03" or about 1 pixel) with the HST's now-defunct Advanced Camera for Surveys High Resolution Camera. Recent HST surveys with a focus on hierarchical triples (Programs 14616 and 15821, PI Porter) have not reported any new detections (or non-detections). Combined with the theoretical expectation for such objects, it seems likely that there are many more hierarchical triples (or quadrupoles) beyond the HST resolution limit which is dozens of primarii radii for typical TNOs. When the tighter binaries are very close together (up to touching ``contact hierarchical triples''), they may be detectable by measuring the unique lightcurve of contact or very close binaries \citep[e.g.][]{2021Icar..35614098S} or multi-chord stellar occultations \citep[][]{2020PSJ.....1...48L}. But beyond quite small separations, such methods are also unable to detect inner binaries. For these reasons, Paper I suggests using non-Keplerian dynamical effects to detect and characterize such objects. 

Whether caused by a non-spherical shape or an unresolved inner binary, non-Keplerian effects on the outer binary are a sensitive probe of the angular momentum distribution \emph{within} TNO binary systems. Such effects are also currently detectable for several TNO binaries as shown in \citet{proudfoot2024beyond} (hereafter Paper II). We thus propose that the strength of non-Keplerian effects can be a probe and constraint on the next generation of SI and gravitational collapse models, in addition to providing information on the shapes and spins of TNO binary components. In this work, we go beyond point mass models for the TNO (148780) Altjira (2001 UQ$_{18}$) to show that it is best explained as an unresolved hierarchical triple. 

The second component of Altjira -- the unnamed S/2007 (148780) 1 -- is only slightly fainter on average, thus this system appears to be a near equal-mass binary. The Keplerian orbit was determined by \citet{Grundy11} to have a period of about 139.5 days, a semi-major axis of about 9904km, and a system mass of about 3.95 $\times$ 10$^{18}$ kg. The orbit was eccentric and inclined by about 35 degrees \citep{Grundy11}. The overall system absolute magnitude is  $H\simeq6$. An approximate thermal measurement lead to estimated diameters of 246$^{+38}_{-139}$ km and 221$^{+31}_{-125}$ km for the two components \citep{Vilenius2014tnos}.This is based on a weak detection and/or upper limits and involves the assumption of equal-albedo, equal-density spheres, so there is significant statistical and systematic uncertainty. These sizes correspond to a density of 0.30$^{+0.50}_{-0.14}$ g cm$^{-3}$ and albedo of 0.04$^{+0.18}_{-0.01}$, suggesting that they could be overestimated. 

\citet{sheppard07} obtained some sparse information on the lightcurve for Altjira and found some discernible variations over 2 nights of observations at the $\sim$0.1 magnitude level and concluded that more data were needed. \citet{Grundy11} reports that the secondary ranges from being 0.3 magnitudes brighter to 0.6 fainter than the primary on 5 epochs spanning a year (and we confirm significant variability below). We thus expect that one or both components to be highly elongated. Long-term non-Keplerian monitoring and/or resolved photometry can better determine how each of the components contributes to the angular momentum budget, but for our present modeling effort we assume that only the primary is a non-spherical quadrupole while the secondary is assumed to be a point-mass. See Paper I for additional discussion of the degeneracies associated with non-Keplerian modeling.

Altjira's heliocentric orbit is near the borderline between the cold and hot classical population, with a free/proper inclination of 5.4$^{\circ}$ \citep{2022ApJS..259...54H}. Depending on the cold/hot cutoff inclination, Altjira has sometimes been classified as part of either population in previous analyses. The free inclination suggests that Altjira is outside the core of the cold classical region, but in actuality the cold and hot population inclinations are best described by von Mises-Fisher distributions (similar to Rayleigh distributions for low dispersion, which themselves are similar to the ``sin i times Gaussian'' model) of the orbit normals \citep{2023MNRAS.522.3298M}.  Thus, some fraction of relatively high free inclination objects are consistent with cold classicals. Using the mixture model of \citet{2023RNAAS...7..143M} for the free inclination distributin of observed non-resonant classical TNOs, we find that $\sim$7\% of TNOs with free inclinations of Altjira are consistent with cold classicals. Furthermore, the color and wide binary nature are better matches for the cold classical population. We propose that Altjira is a cold classical with an inclination that is higher than usual.

\citet{Grundy11} find that Altjira is predicted to have mutual events -- where the primary and secondary shadow and/or occult one another -- around 2028. We confirm and update this prediction below. Although the observation of such mutual events is challenging and the interpretation even more so \citep[e.g.,][]{Rabinowitz2020complex}, it provides an exciting opportunity to probe the properties of the components of Altjira without having to wait for a stellar occultation. 

In 2023, HST observed Altjira to get two more precise astrometric measurements as part of Program 17206 (PI Proudfoot). We combine these observations, a new Keck observation, and previously published observations as described in Section \ref{sec:obs}. For the new HST data, we measure the mutual astrometry and photometry of these objects using a new open-source fitting program called \texttt{nPSF} located at \url{https://github.com/dragozzine/nPSF} which is described in Section \ref{sec:npsf}. We then use the non-Keplerian astrometric orbit fitter \texttt{MultiMoon} (Paper I) to fit the new Altjira data in Section \ref{sec:methods}. In addition to the usual Keplerian model, we use a binary quadrupole--point-mass model and a hierarchical triple model to describe the observations. We find that a close hierarchical triple is the preferred model for Altjira in Section \ref{sec:results}. We provide an ephemeris for Altjira to support the ongoing mutual event season, including quantifying the importance of using a non-Keplerian model, in Section \ref{sec:mutual}. We interpret our results in Section \ref{sec:discussion} and conclude in Section \ref{sec:conclusions}.


\section{Observations} \label{sec:obs}

The observations for determining Altjira's binary orbit focus on precise relative astrometric positions. We use the original 8 data points from \citet{Grundy11} as they are. These come from HST/ACS, HST/WFPC2, and Keck as discussed in \citet{Grundy11}. We combine these with an additional observation obtained from Keck and two new observations from HST/WFC3. 

Keck observations of the system were obtained on 2020 January 5 at a mean time of 9:06 UTC and mean airmass of 1.34.  We used the NIRC2 narrow field infrared camera with a plate scale of 100.5 pixels per arcsec along with the Keck 2 laser guide star adaptive optics system \citep[LGS AO][]{2006SPIE.6272E..01L} to correct for the smearing effect of atmospheric turbulence.  A nearby 13th mag star was used for tip-tilt correction.  A total of six usable 180 second exposures were recorded through the H filter, with exposure pairs recorded in each of three consecutive dither positions.  After flat fielding and subtracting the other dither positions, astrometry was done on each of the co-added pairs by fitting a Lorentzian pointspread function with the same width parameters to each of the two components in the image.  Plate scale and camera orientation details were taken from \citet{2010ApJ...725..331Y}.  Based on prior experience with this instrument configuration with targets of similar faintness, the 1-sigma astrometric uncertainties were taken as 3 mas, although the scatter between the measurements from the 3 dither positions was somewhat smaller than that. 

As part of HST Program 17206, two additional single-orbit visits of Altjira were acquired in 2023. These visits each consisted of 6 dithered observations using HST's WFC3 scheduled at times at which our preliminary analysis indicated that the position of the two system components had the most uncertainty in previous modeling (i.e., high ``information gain'' as discussed in Paper II). These two visits were designed to achieve high astrometric precision by maximizing signal-to-noise, with observations using HST's wide F350LP filter.

\section{Fitting an arbitrary number of Point Spread Functions with \texttt{nPSF}} \label{sec:npsf}

We here introduce a new open-source (\url{https://github.com/dragozzine/nPSF}) precise Point Spread Function (PSF) fitting routine called \texttt{nPSF}. \texttt{nPSF} works similar to the description of PSF-fitting routines in \citet{2009AJ....137.4766R} and \citet{2008Icar..197..260G} which are standard in the field. We then use \texttt{nPSF} to measure astrometry for the two observations from HST/WFC3. 

\subsection{How \texttt{nPSF} works}

For solar system small bodies, it is common that the separation of the sources is several pixels and the width of the PSF (e.g., the Full Width at Half Maximum or FWHM) is a few pixels. Occasionally, it is possible to perform relative astrometry with close sources, but when the separation is less than $\sim$2 pixels, it becomes very challenging, if not impossible, even for equally bright sources \citep[e.g.][]{benecchi201047171}. Even when the separation is clearly larger than the FWHM (the components are ``resolved''), the light from the sources may overlap, requiring simultaneous modeling of each of the sources. 

Precise relative astrometry depends on accurately measuring the center of light of two or more sources. 
HST benefits from having a very stable and well-understood PSF as a function of camera, filter, approximate pixel location, and telescope focus value. We use theoretical supersampled PSFs as provided by the well-known TinyTim software package \citep{2011SPIE.8127E..0JK} with occasional approximations (e.g., not using the exact pixel locations of the sources in the PSF generation process). Though TinyTim is not officially recommended for accurate WFC3 PSF modeling, we note that precise relative astrometry does not require highly accurate PSFs because measuring the center of light (basically the weighted average position) is relatively robust relative to the statistical precision for these faint objects.

\texttt{nPSF} models each image separately. A subsection of the image (e.g., 50 x 50 pixels) approximately centered on the primary component is selected. Cosmic rays are cleaned using the method of \citet{van2001cosmic} as implemented by \citet{mccully2019astro}. The median of the pixels is subtracted to remove any background. Pixel-by-pixel uncertainties are accounted for using a Poisson noise model. This allows us to gather uncertainties from each image individually which are fully propagated to astrometric uncertainties. 

Using standard methods, a model image is constructed for comparison to the observations. At the beginning of the analysis, the user can choose $n$, the number of PSFs to include in the model. The model requires multiple free parameters, primarily the $x$ and $y$ positions and ``heights'' (e.g., the total brightness) for each PSF. We also allow the HST telescope focus to be a free parameter by generating a grid of model PSFs with different focus values. The floating focus parameter is then rounded to the value of the nearest pregenerated PSF. With a fine enough grid of focus values, we closely approximate having a fully floating focus parameter. This approximately models HST ``breathing'' and also allows for some compensation of the fact that the theoretical model PSF is likely not an exact match to the observed PSF. The PSFs are inserted at their model locations, using fractional pixel shifts of the supersampled theoretical PSFs. We include the charge diffusion kernel contained within each modeled TinyTim PSF. Using our noise model, we calculate the overall likelihood by summing the log likelihoods of each individual pixel; that is, we assume all pixels are independent. 

We use Bayesian parameter inference to explore the range of models (although an optimization mode is also implemented). This determines the ``posterior'' probability distribution, e.g., a list of self-consistent sets of parameters that are a good match to the data and which also automatically provide uncertainties and correlations between each parameter. We use the \texttt{emcee} package to perform the posterior sampling using Affine-Invariant Ensemble Sampling via Markov Chain Monte Carlo \citep{foreman2013emcee,Foreman-Mackey2019}. Priors are generally uninformative and initial guesses are set by visual inspection of the image. After fitting for image-coordinates, the \texttt{pixel-to-world} function in \texttt{astropy} \citep{astropy} uses information in the image headers to calculate the posterior distribution of the relative astrometric offsets in Right Ascension and Declination. Internally, \texttt{MultiMoon} uses time as measured from a clock located at the TNO and relative astrometry in ecliptic coordinates. A final step in \texttt{nPSF} outputs a single line of astrometric data in these units that can be copied directly into an observation file for \texttt{MultiMoon}. 

ZZZ Write a short paragraph about Bayesian parameter inference with MCMC. 

By design, the fitting procedure of \texttt{nPSF} is very similar to that used for \texttt{MultiMoon} (see Paper I) including a pre-specified ``burn in'', which is a clustering algorithm to remove under performing "walkers", (\textbf{strings of multiprocessor calculations exploring the space,}) and plots generated for the final set of parameters to allow for visual inspection of performance metrics like convergence. Unlike \texttt{MultiMoon}, the time taken for a single likelihood evaluation is very short (tens of milliseconds), so \texttt{nPSF} is not parallelized. A typical run takes less than $\sim$1 day of wall clock time for a single CPU to analyze a single image even for an extensive exploration of parameter space. 

We performed a suite of injection and recovery tests to confirm that \texttt{nPSF} was working well and to test the limits of its performance. For example, under ideal injection and recovery conditions, we find that it can just successfully recover a secondary when the separation in pixels is about the same as the brightness difference in magnitudes. We have also validated \texttt{nPSF} on a variety of past HST programs, focusing on TNB astrometry and find adequate consistency with published observations. 

We note that even though the processes are similar in theory, we found that PSF-fitting analyses can be very different in practice depending on where they fell along three axes: semi-resolved vs. well-resolved, similar brightness vs. unequal brightnesses, and measurement of a known source vs. detecting a new source (or placing upper limits). \texttt{nPSF} does well for known objects with similar brightnesses even with close separations. It is not ideal for semi-resolved unequal brightnesses without extra care (e.g., setting a prior for the brightness ratio of the components). It can be used to provide approximate upper limits for non-detections, but injection/recovery tests would be more accurate and significant. 

\subsection{Application of \texttt{nPSF} to Altjira HST images}

We used \texttt{nPSF} on the two epochs (Feb and Oct 2023) of HST/WFC3 images of Altjira. All 12 images were run with initial guesses based on inspection of the image and all model parameters for a 2-PSF fit. We used 2500 burn in steps and 2500 convergence steps for 100 walkers, except for one image from the October visit that needed 500 extra steps to converge. One image from each visit was excluded due to systematic errors and problems with cosmic ray masking, which didn't allow their fits to converge. The other five images appeared converged based on visual inspection and had no issues. The derived astrometry and uncertainties is shown in Table \ref{tab:npsfresults}.

\begin{deluxetable*}{ccccccc}
    \tablecaption{Derived Astrometry From HST images taken in February and October 2023}
    \tablewidth{0pt}
    \tablehead{
    Time &  $\Delta$ Lat. &  $\Delta$ Long. &  $\Delta$ Lat. Error &  $\Delta$ Long. Error}
    \startdata
    \\\hline
        2459982.70877 & -0.067317 & 0.157316 & 0.00091621 & 0.00080187  \\
        2459982.71397 & -0.066984 & 0.159926 & 0.00082586 & 0.00076244  \\
        2459982.71927 & -0.067997 & 0.158797 & 0.00090732 & 0.00081523  \\
        2459982.72456 & -0.065999 & 0.158752 & 0.00092021 & 0.00082179  \\
        2459982.72976 & -0.067761 & 0.158984 & 0.00088429 & 0.00079804  \\
        \\\hline
        2459982.71927 & -0.067212 & 0.158755 & 0.00089078 & 0.00079987 \\
        \\\hline
        2460240.8251 & 0.083471 & -0.169464 & 0.00110204 & 0.00102173  \\
        2460240.8303 & 0.084002 & -0.170261 & 0.00110006 & 0.00102710  \\
        2460240.8406 & 0.082228 & -0.171161 & 0.00113814 & 0.00108419  \\
        2460240.8457 & 0.083687 & -0.170873 & 0.00110498 & 0.00108649  \\
        2460240.8509 & 0.081580 & -0.171009 & 0.00115472 & 0.00109073  \\
        \\\hline
        2460240.8418 & 0.0829938 & -0.17055 & 0.00111999 & 0.00106205 \\
    \enddata
    \tablecomments{Table of derived astrometry from each new HST/WFC3 image from \texttt{nPSF} as well as the averages from each visit. Time is in ``System'' Julian Date, e.g., with the light-travel time correction to the TNO included. Relative positions are given in J2000 ecliptic coordinates.  }
    \label{tab:npsfresults}
\end{deluxetable*}

The results for Altjira have the observed positions reasonably near the predicted positions and have very small (sub-milliarcsecond) uncertainties. In February and October, the secondary was 0.21 and 0.57 magnitudes fainter than the primary, repsectively. 

\texttt{nPSF} uses pixel-by-pixel uncertainties that are used to determine confidence intervals for relative positions for each individual image. Current practice combines these all into a single astrometric measurement by taking the mean and standard deviation of the best-fit positions from each image (after discarding obvious outliers). The rationale for this is that such a technique is better at incorporating systematic uncertainties, though using the standard deviation of best fits is not an ideal way of propagating uncertainty, especially when only a few images are used. For these observations, the standard deviation of best fit values is quite similar to the reported uncertainties. If the image-by-image uncertainties were accurate, the standard deviation of the positions would be about $\sqrt{5}$ times smaller; this suggests that there are systematic effects so that our image-by-image uncertainties are optimistic. 

We explicitly tested the difference between using image-by-image uncertainties and the best-fit averaging method in our \texttt{MultiMoon} fits and found that the results were not significantly different in accuracy or precision. For consistency with previous modeling methods, and to account for systematic uncertainties, we use the average points and their standard deviations in our detailed analysis below. 

Most of the time, the relative motion of the components of Altjira is less than 1 milliarcsecond per hour. Thus, we do not expect to see, nor do we see, motion during the course of the observations. In other solar system small bodies, motion can be much faster suggesting that image-by-image \texttt{nPSF} analysis would be more accurate in providing astrometry that includes information about the motion instead of averaging it out. 

We discuss below the possibility that one or both of the components of Altjira may themselves be a close binary. We checked the residuals of the \texttt{nPSF} astrometry fits but found no clear evidence for a third object in the images. This is consistent with the expected separations of these binaries derived below which are far below the resolution limit. 

\section{Methods} \label{sec:methods}

Combining all the relative astrometric measurements for Altjira provides the observational constraints on our orbit modeling. These are summarized in Table \ref{tab:obs}. 

\begin{deluxetable*}{ccccCCCC}
\tablecaption{Relative Astrometry for 148780 Altjira}
\tablewidth{0pt}
\tablehead{
Julian Date & Date & Telescope & Instrument/Camera & \Delta \alpha \cos{\delta} & \Delta \delta & \sigma_{\Delta \alpha \cos{\delta}} & \sigma_{\Delta \delta}}
\startdata
2453953.767 & 2006 Aug 6 & HST & ACS-HRC & -0.17182 & 0.06302 & 0.00109 & 0.00408 \\
2454306.58 & 2007 Jul 25 & HST & WFPC2-PC & -0.16332 & -0.03667 & 0.00167 & 0.00168 \\
2454320.431 & 2007 Aug 8 & HST & WFPC2-PC & -0.27679 & -0.00742 & 0.00204 & 0.001 \\
2454380.396 & 2007 Oct 6 & HST & WFPC2-PC & -0.05278 & 0.04552 & 0.001 & 0.0019 \\
2454416.885 & 2007 Nov 12 & HST & WFPC2-PC & 0.14054 & -0.06317 & 0.00138 & 0.00118 \\
2454672.799 & 2008 Jul 25 & HST & WFPC2-PC & 0.15723 & -0.00858 & 0.00133 & 0.00143 \\
2455176.892 & 2009 Dec 11 & Keck 2 & NIRC2 & -0.36534 & 0.03793 & 0.003 & 0.003 \\
2455412.109 & 2010 Aug 3 & Keck 2 & NIRC2 & -0.03594 & -0.03938 & 0.006 & 0.003 \\
2458884.879 & 2020 Feb 5 & Keck 2 & NIRC2 & 0.18001 & -0.06656 & 0.003 & 0.003 \\ 
2459982.981 & 2023 Feb 7 & HST & WFC3 & 0.16501 & -0.04991 & 0.00092 & 0.00079 \\
2460241.099 & 2023 Oct 23 & HST & WFC3 & -0.17699 & 0.06817 & 0.00065 & 0.0011 \\
\enddata
\tablecomments{The relative right ascension ($\Delta \alpha \cos{\delta}$) and declination ($\Delta \delta$) positions of Altjira's secondary, along with measurement uncertainties in arcseconds. Julian Date is geocentric; the specific location of the observatories is negligible and ignored. \label{tab:obs}}
\end{deluxetable*}

We fit these data using three different orbit models: a Keplerian (two point-mass) model, a binary non-Keplerian model of a quadrupole and a point mass, and a non-Keplerian model of three point masses. To accommodate non-Keplerian orbital modeling, we use the \texttt{MultiMoon} code powered by the \texttt{SPINNY} $n$-quadrupole integrator. A detailed explanation and validation of \texttt{SPINNY} and \texttt{MultiMoon} is provided in Papers I and II, but we provide an overview of key aspects here before discussing our analysis for Altjira specifically.

\subsection{The \texttt{SPINNY} $n$-quadrupole model}

\texttt{SPINNY} integrates the orbit and spin dynamics of an arbitrary number of quadrupoles or point-masses \citep[as in][]{correia2018chaotic}. Quadrupoles are characterized by a spherical harmonic expansion of the potential where $J_2R^2$ and $C_{22}R^2$ quantify the oblateness and prolateness respectively. The orientation of the quadrupole is described using Euler angles equivalent to the orbital elements such as the ``spin inclination'' $i_{sp}$; these are defined relative to the J2000 ecliptic frame. For small bodies like the components of Altjira, the dominant contribution to the quadrupole strength is the shape and orientation of the body, not the interior mass distribution. However, these spherical harmonic terms do not uniquely provide information on the shape, so an assumed shape model is required to calculate a specific value of $J_2$. For example, assuming a triaxial ellipsoid model, $J_2R^2$ and $C_{22}R^2$ can be used to determine the overall shape. See Papers I and II for more detailed discussion. 

\texttt{SPINNY} calculates the spin-orbit coupling (including the back torque of the orbit on the spin of the body) which leads to both apsidal and nodal precession. In practice, there are many degeneracies between model parameters and some parameters that are not constrained by the observations. For example, the spin rate, the prolateness ($C_{22}R^2$), and the angle associated with the long axis of the quadrupole ($\omega_{sp}$) are practically unconstrained when the spin period is much shorter than the orbital period because the rotational effects easily average out. In the case of Altjira, short-term variability from \citet{sheppard07} combined with the relatively wide separation of the binary means that these parameters are not expected to be meaningfully constrained, though we do include them in our model for completeness. Furthermore, we have found that there can be many degeneracies when considering a model with two quadrupoles, even if only the oblateness terms are considered. Generally speaking, the observed non-Keplerian motion is mostly due to orbital precession which is the sum of precessions caused by the two quadrupoles. (Technically, this assumes that most of the angular momentum is in the orbit, which is the case for Altjira; the case where there are three angular momenta reservoirs is poorly understood.) Similar to how a Keplerian model can only determine the sum of the masses, Altjira's observational data can only constrain a weighted sum of the $J_2R^2$ values for the two components. Based on theoretical expectations (see Paper I and Paper II), we hypothesize that the approximate quantity measured is $M_1J_{2,1}R_1^2 \cos \phi_1 + M_2J_{2,2}R_2^2 \cos \phi_2$, where $M$ is the mass and $\phi$ is the obliquity of the spin to the orbit. This also means that the unknown mass partitioning between the two components affects the interpretation of the shapes; on the other hand, extensive observations and non-Keplerian modeling can break the mass degeneracy in certain cases. 

Thus, for simplicity, we consider only the primary as having a non-spherical shape and thus call this the binary quadrupole--point-mass model. In actuality, the two components are similar brightnesses and thus likely have similar values of $J_2R^2$, suggesting that the two components individually might each have about half of the $J_2R^2$ predicted in our model. 

At a sufficiently large separation, a close binary appears like a large quadrupole in terms of its dynamical effects. For example, \citet{proudfoot2024beyond} found that a close-in satellite would have:
\begin{equation}\label{eqn:j2binary}
    J_{2}R^2 = \frac{1}{2} \frac{q}{\left(1+q\right)^2} a_s^2
\end{equation}
\begin{equation}\label{eqn:c22binary}
    C_{22}R^2 = \frac{1}{4} \frac{q}{\left(1+q\right)^2} a_s^2
\end{equation}
\noindent where $q = m_{Aa}/m_{Ab}$, $m_{Aa}$ and $m_{Ab}$ are the masses of the masses of the two (unresolved) components and $a_s$ is the semi-major axis of the secondary's orbit around the barycenter. For an equal mass binary composed of two equal size spheres, $J_2R^2$ = $\frac{1}{2} a_{A}^2$ where $a_{A}$ is the separation between the two components (twice the barycentric $a_s$). 

As discussed below, the inferred value of $J_2R^2$ for the components of Altjira suggests such an elongated shape that a close inner binary is a reasonable hypothesis. We thus consider also a (close) hierarchical triple model composed of three point masses, also provided by \texttt{SPINNY}. Again, to reduce model complexity and avoid degeneracies, our model assumes that the primary is composed of an unresolved close binary (of two point masses). This ``inner'' binary is orbited, as before, by the observed second component in the system which constitutes the ``outer'' binary. 

Many other configurations are possible such as two ``inner'' binaries in a hierarchical quadruple system or that the secondary is a close binary orbiting by a triaxial primary. The differences between these models cannot be probed by the current observational data, though precise lightcurve observations would help. However, a statistically significant detection of $J_2R^2$ in our simplified model requires that the two components together deviate from point masses. It also provides measurements on the sum of the angular momentum of the two components and is thus an observational constraint of how the system formed by understanding how angular momentum is partitioned in the system as discussed above. 

\bp{In principle, \texttt{SPINNY} can also model the Sun's gravitational influence on a binary (or triple) system. However, within a critical semi-major axis $a_{crit}$ \citep[given by equation 3 of][]{nicholson2008irregular}, a binary's dynamics are dominated by $J_2$ rather than the Sun's influence. For the mass and $J_2$ of Altjira, $a_{crit} \sim$ 30,000 km. Since Altjira's semi-major axis is well within this boundary, solar gravity can be safely ignored. Thus for all modeling performed here, we neglect the Sun's influence.}

\subsubsection{Bayesian parameter inference from \texttt{MultiMoon}}

Given a set of model parameters, \texttt{SPINNY} efficiently calculates the relative positions of all the components at all the observation times. As usual, these relative positions are then projected into the plane of the sky as seen from the geocenter (the relative position of the observatories is negligible and ignored), including the light-travel time correction. This provides a model for relative astrometric positions that is then compared to the data assuming independent Gaussian noise with uncertainties listed in Table \ref{tab:obs}. That is, the $\chi^2$ goodness-of-fit metric for these model parameters is calculated in the usual way. For Altjira, a single $\chi^2$ calculation takes tens of milliseconds for a Keplerian model (which is performed analytically without using \texttt{SPINNY}) and about a second for the non-Keplerian models. 

To explore and understand how the fit is improved by varying parameters, \texttt{MultiMoon} uses Bayesian parameter inference powered by \emph{emcee} (Foreman-Mackey et al. 2013) which uses Affine-Invariant Ensemble Sampling Markov Chain Monte Carlo. Details for how \texttt{MultiMoon} works are provided in Paper I. For each \texttt{MultiMoon} run, various parameters can be ``fixed'', or frozen, while other parameters ``float''. Initial guesses are provided by a large number of ``walkers'' which allows for multiprocessor calculations that take advantage of our supercomputer resources. After completing a ``burn-in'' -- during which the walkers identify the region of the best fit -- under-performing walkers are removed using a clustering analysis. At this point, the now-converged walkers continue to explore the parameter space where each walker-step provides a sample from the ``posterior'' probability distribution of the parameters. The ensemble of these samples can then be used to calculate which parameters provide adequate fits to the data. For example, we report the median (50th percentile) and uncertainties based on confidence intervals (16th and 84th percentiles) that are chosen to be comparable to Gaussian ``1-$\sigma$'' values since most posterior distributions are nearly Gaussian. Both the ``burn in'' and total run length are set in advance, but confirmed to be appropriate by visual inspection of the chains using plots provided by \texttt{MultiMoon}.

Calculation of the Bayesian posterior probabilities requires the choice of a ``priors'' for all the parameters and a specific noise model that determines the ``likelihood'' or probability of obtaining the observations given an assumed forward model. The likelihood assumes Gaussian uncertainties, e.g., the log likelihood (to an unimportant constant) is given by $\log \mathcal{L} = -\frac{1}{2} \chi^2$ \citep[e.g.][]{2010arXiv1008.4686H}. We generally choose ``uninformative'' flat priors in all our parameters to allow the observational constraints to control the posterior probabilities. There are two major exceptions. We restrict $C_{22}R^2 \le \frac{1}{2} J_2R^2$ as physically reasonable (see discussion in Paper I and \citet{correia2018chaotic}). When modeling a hierarchical triple, we also require that the secondary be less massive than the primary, where this is now referring to the two (unresolved) components in the ``inner'' binary. We believe that these prior choices do not significantly affect our conclusions. 

\subsection{Binary Non-Keplerian Runs}

We now turn specifically to our analysis of Altjira. The binary non-Keplerian run refers to our model where the primary is a quadrupole and the secondary is a point mass. Excluding a variety of preliminary and exploratory models (including those discussed in Paper II), our final model of this system used \texttt{MultiMoon} with 960 walkers, 20,000 burn-in steps, and 20,000 posterior sampling steps. The resulting 19.2 million posterior samples provide a highly accurate sampling. The \texttt{SPINNY} integration tolerance was $10^{-10}$ consistent with other analyses and tested to be sufficiently accurate to not introduce significant systematic uncertainties. We fix the spin rate parameter and approximate radius to 1.7453 $\times$ $10^{-4}$ (rad/s) and 123 km and confirmed that these fixed parameters do not affect the results. (These parameters affect how the orbital motion changes the unobserved spins.)

\subsection{Hierarchical Triple Runs}

As above, the three-point-mass Altjira fits used 960 walkers, 20,000 burn-in steps and 20,000 sampling steps. We confirmed that a \texttt{SPINNY} tolerance of $10^{-9}$ did not affect the results. 

Three-point-mass models can have 15 free parameters (2 sets of 6 primary-centric orbital elements and 3 masses) and without astrometric measurements of the inner binary, there are few constraints on its orbit. As a result, our goal was not to provide a full posterior distribution for these fits, but rather to find examples of parameter sets that have high posterior probability. This provides a ``proof of concept'' for three-point-mass models that can adequately match the data. Although it is not an optimizer (see \citet{2018ApJS..236...11H} for discussion), we use \texttt{MultiMoon} to explore the parameter space, usually assuming circular orbits. The parameters were initialized by taking the $J_2R^2$ from the non-Keplerian binary fits and assuming this was provided by an equal-mass inner binary using Equation \ref{eqn:j2binary} for the separation and using the orientation angles (e.g., $i_{sp}$ and $\Omega_{sp}$) for the orientation of the inner binary. We note  that we did \emph{not} include the center-of-mass--center-of-light offset for the unresolved inner binary which is a small correction for our analysis, but should be considered in future work. 

\begin{figure}[p]
\centering
\includegraphics[width=1.0\textwidth]{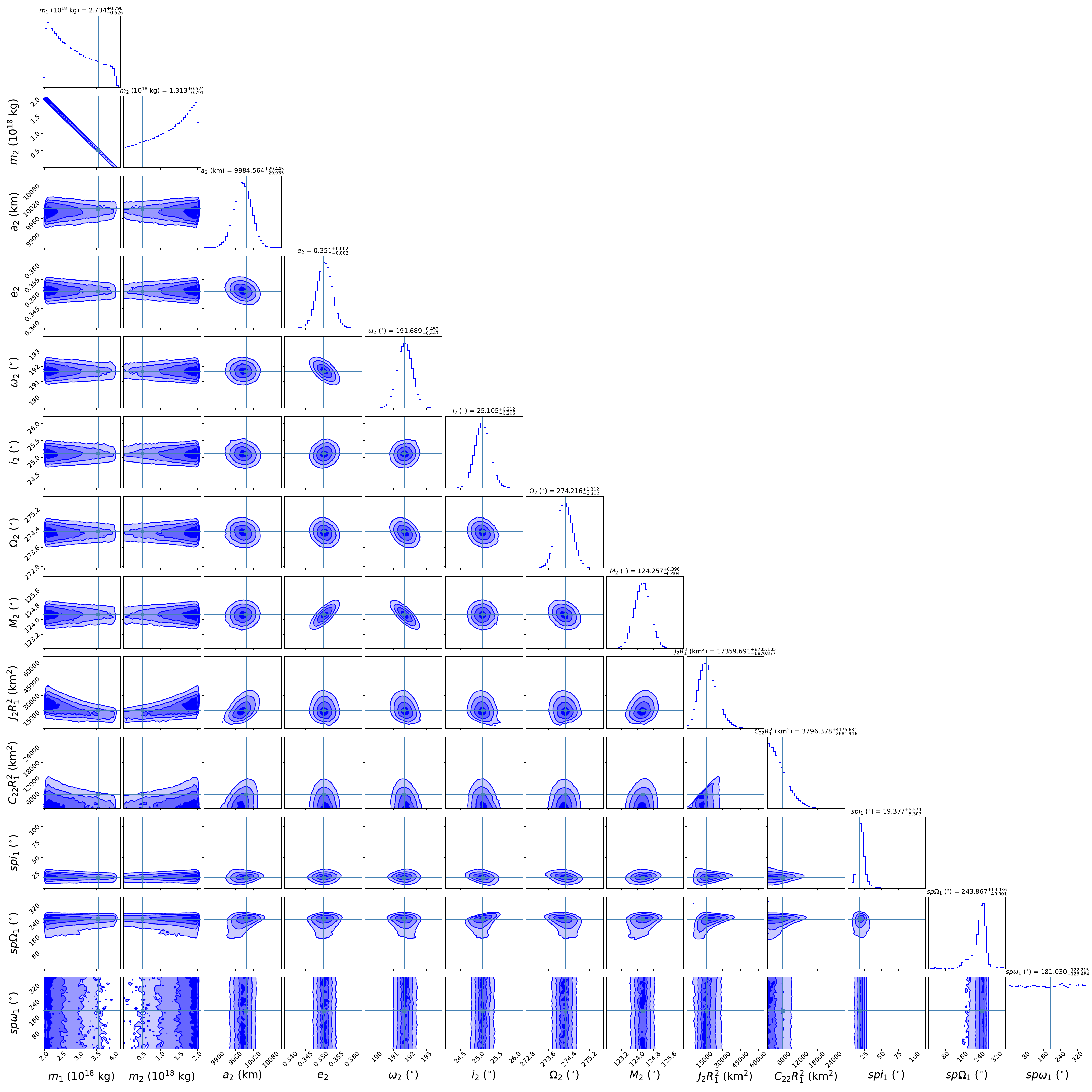}
\caption{\label{fig:corner} Corner Plot for the binary Quadrupole–point-mass non-Keplerian Altjira orbit fit. Along the tops of the columns are the marginal posterior distribution for each parameter, which can be used to determine the best fit and uncertainty for Gaussian distributions. The contour plots shows the 1,2, and 3$\sigma$ levels of the joint posterior distributions between every pair of given parameters. Of particular interest is the J2R2 posterior which strongly disfavors the Keplerian model ($J_2R^2 = 0$). All angles are measured relative to the the J2000 ecliptic plane on JD 2454300.0 (2007 July 18th, 12:00 UT).}
\end{figure}

\begin{deluxetable*}{lCCCC}
\tablecaption{Non-Keplerian Orbit Solution for 148780 Altjira}
\tablewidth{0pt}
\tablehead{
\colhead{Parameter} & \colhead{} & \colhead{Posterior Distribution} & \colhead{Best Fit} &\colhead{\textbf{Keplerian Fit}}
}
\startdata
Fitted Elements & & & \\
\qquad Primary mass ($10^{18}$ kg) & M_1 & 2.7335^{+0.79}_{-0.52}* & 3.547* & \textbf{3.841}\\
\qquad Secondary mass ($10^{18}$ kg) & M_2 & 1.3125^{+0.52}_{-0.80}* & 0.5114* & \textbf{1.537}\\
\qquad Semi-major axis (km) & a & 9945.56^{+30}_{-30} & 9995.99 & \textbf{9944}\\
\qquad Eccentricity & e & 0.3511^{+0.0025}_{-0.0024} & 0.3508 & \textbf{0.3523}\\
\qquad Inclination ($\degr$) & i & 25.105^{+0.21}_{-0.21} & 25.1134 & \textbf{25.368}\\
\qquad Argument of periapse ($\degr$) & \omega & 191.69^{+0.45}_{-0.45} & 191.65 & \textbf{192.0} \\
\qquad Longitude of the ascending node ($\degr$) & \Omega & 274.21^{+0.31}_{-0.31} & 274.26 & \textbf{274.0}\\
\qquad Mean anomaly ($\degr$) & \mathcal{M} & 124.25^{+0.40}_{-0.41} & 124.28 & \textbf{124.2}\\
\qquad Primary zonal gravitational harmonic & J_2R^2 & 17359^{+8759}_{-6906} & 16219 &\\
\qquad Primary sectoral gravitational harmonic & C_{22}R^2 & 3796^{+4203}_{-2691}* & 5419* &\\
\qquad Primary rotation axis obliquity ($\degr$) & i_{sp} & 19.377^{+5.61}_{-5.34} & 17.614 &\\
\qquad Primary rotation axis precession ($\degr$) & \Omega_{sp} & 243.86^{+19}_{-40} & 249 &\\
\qquad Primary rotation axis position ($\degr$) & \omega_{sp} & 181.03^{+123}_{-124}* & 191.31* &\\
Derived parameters & & & \\
\qquad System mass ($10^{18}$ kg) & M_{sys} & 4.0453^{+0.0356}_{-0.0360} & 4.0588 & \textbf{5.38}\\
\qquad System density (g cm$^{-3}$) & \rho_{sys} & 0.37^{+0.74}_{-0.24} & 0.30 &\\
\qquad Primary obliquity w.r.t. orbit ($\degr$) & \phi & 14.44^{+12.23}_{-6.20} & 11.810 &\\
\qquad Non-Keplerian orbital period (days) & P_{orb} & 139.68^{+0.013}_{-0.014} & 139.68 &\\
\enddata
\tablecomments{Orbital and Physical parameters from our non-Keplerian binary quadrupole--point-mass fit for Altjira. Parameters marked with (*) are not well-determined. Full posterior distributions of parameters and degeneracies can be seen in Figure \ref{fig:corner}. For example, although nominally a non-Keplerian fit can break the mass degeneracy between the two objects, there are not enough data in this case to do so reliably; thus, only the sum of the masses is meaningfully constrained (and unchanged from the Keplerian fit of \citet{Grundy11}). All fitted angles are relative to the J2000 ecliptic plane on Altjira-centric JD 2454300.0 (2007 July 18th, 12:00 UT). For the system density, we use diameters of 246$^{+38}_{-139}$ km and 221$^{+31}_{-125}$ km \citep{Vilenius2014tnos}, though these are likely overestimates. This fit has $\chi^2$ of 14.6 and the probability that a $\chi^2$ value this high would be due to random noise is 8.6\%, so that the fit is statistically acceptable. The non-Keplerian fit is a statistically significant improvement over the Keplerian fit based on the improved $\chi^2$. Some hierarchical triple three-point-mass models provide even better fits to the observations, but these were not explored rigorously as discussed in the text. 
\label{tab:orbitfit}}
\end{deluxetable*}

\section{Results} \label{sec:results}

Using an exploratory analysis by \texttt{MultiMoon} on less data, Paper II found that Altjira showed evidence for non-Keplerian motion. The observed large changes in brightness also require a model that moves beyond point-mass spheres. We confirm statistically significant non-Keplerian motion in our more detailed analysis that includes the three new observations. 

First, we consider a Keplerian model. This model reaches a maximum log likelihood of 14.7 corresponding to a minimum $\chi^2$ of 29.4 and a reduced $\chi^2$ of 2.09. Using the $\chi^2$ statistical distribution, there is only a 0.97\% chance that the residuals from the Keplerian model are consistent with the assumed noise model. As in Paper II, the orbital parameters for our new Keplerian model were very similar to the previously published parameters in \citet{Grundy11} and to the orbital parameters in the non-Keplerian fits. 

Note that non-Keplerian models explicitly include the Keplerian model as a subset; for example, the binary non-Keplerian model would return $J_2R^2 \approx 0$ if the true model was Keplerian. For this reason, and because a non-Keplerian model is a more accurate representation of the real Altjira system, we prefer to think of the Keplerian model as an extreme version of the non-Keplerian model that has all quadrupole components set to 0 as a prior assumption. 

The parameters (giving the 16th, 50th, and 84th percentiles from our posterior samples) for the binary non-Keplerian model of Altjira are reported in Table \ref{tab:orbitfit}. The posterior distribution for all the parameters is also plotted as a corner plot in Figure \ref{fig:corner} which includes 1-dimensional marginal distributions for each parameter and 2-dimensional joint distributions.

\begin{figure}
\centering
\includegraphics[width=0.8\textwidth]{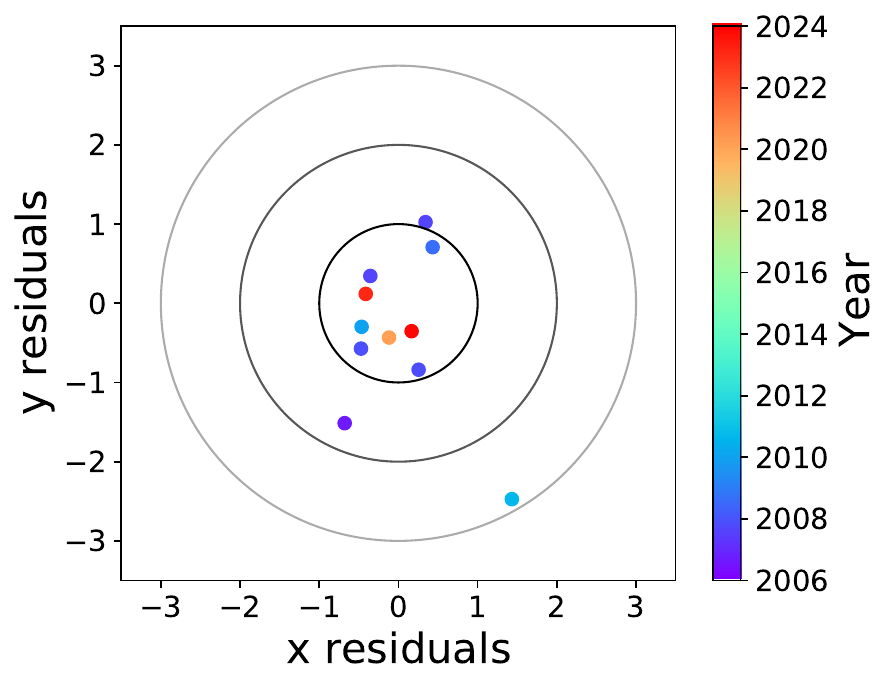}
\caption{\label{fig:residuals} Astrometric residuals for observations shown in Figure \ref{tab:obs} for the binary quadrupole--point-mass fit. Points are colored as a function of time to illustrate that the residuals show no major time-dependence. These residuals are consistent with Gaussian noise (p-value assuming a $\chi^2$ distribution of 8.6\%). }
\end{figure}

The fit is a significant improvement with a log likelihood of -7.6, with a minimum $\chi^2$ of 15.2 and a reduced $\chi^2$ of 1.68. The probability that a $\chi^2$ value this poor would have occurred assuming an accurate noise model is 8.6\%. The best-fit is an excellent match to the data as demonstrated by the residuals plot in Figure \ref{fig:residuals}. The improvement in $\chi^2$ from the Keplerian model ($\Delta \chi^2 = 14.2$) is highly statistically significant, even with the addition of 6 additional parameters. We find a statistically significant $J_2R^2$ which excludes 0 (the Keplerian case) at $\sim$2.5-$\sigma$. While this is not overwhelming evidence for non-Keplerian motion, it is clearly preferred over the Keplerian model. 

As expected, $C_{22}R^2$ and $\omega_{sp}$ are not meaningfully constrained. In theory, non-Keplerian motion can break the Keplerian degeneracy and measure each mass individually, but our data and model are not sufficient and the mass estimates in Table \ref{tab:orbitfit} and Figure \ref{fig:corner} should not be used. We find that the quadrupole orientation ($\phi$) is relatively well-aligned (14$^{+12}_{-6}$ degrees) with the orbit suggesting that the obliquity of the primary (in the quadrupole model) or the orbit pole of the inner binary (in the hierarchical triple model) is well aligned to the outer binary.

The nominal $J_2R^2$ of about 17000$^{+9000}_{-7000}$ km$^2$ is quite high for the apparent size of the primary (as estimated from thermal measurements). Even though this value is really like the sum of the oblateness of the two components, it is still higher than would be expected for equilibrium shapes (e.g., Jacobi ellipsoids). This motivated our investigation of a hierarchical triple three-point-mass model. 

As discussed above, we did not perform an exhaustive search of the hierarchical triple parameter space. Our analyses did identify some degeneracies and possibly even multiple solution modes. Even so, we found an example of an excellent fit to the data with such a model, assuming a circular inner binary orbit. This fit has a log likelihood of -5.9 and a $\chi^2$ of 11.8, a substantial improvement to the quadrupole--point-mass model, even considering the additional two free parameters. This model has near-equal masses for the inner two components (with a total mass of 3 $\times$ 10$^{18}$ kg) and a (primaricentric) semi-major axis of 124 km (corresponding to an orbital period of about 5.5 hours). We note that this is closer than the sum of the proposed radii, though it is similar to a contact configuration using the 1-$\sigma$ sizes. This is somewhat smaller than expected as it implies an effective $J_2R^2 \approx 7700$ km$^2$ from Equation \ref{eqn:j2binary}. It is possible that a configuration with a semi-major axis around the expected 184 km separation is also consistent with the observations. The inclination and longitude of ascending node of the inner binary for this model are 42$^{\circ}$ and 190$^{\circ}$, respectively, relatively similar to that expected from the binary non-Keplerian fit. Due to the various degeneracies and limited data, we consider the quadrupole--point-mass and the hierarchical triple fit to be reasonably self-consistent and both adequate descriptions of the data.

We discuss the implications and interpretation of these fits in Section \ref{sec:discussion} below. 

\section{The Ongoing Altjira Mutual Event Season} \label{sec:mutual}

\citet{Grundy11} noted that Altjira would have a mutual event season (e.g., shadowings and occultations of the two components) centered around 2028. With our new updated orbit fits, we confirm and update this prediction. 

The shapes of the components of Altjira are non known, but are clearly not spherical based no our measurement of $J_2R^2$ as well as significant ($\sim$1 magnitude) variability in the relative brightness of the components. Including detailed lightcurve observations could provide additional insight to the shapes, but only in the presence of multiple assumptions like uniform surfaces, equal albedos, and equal densities. Stellar occultations would provide valuable unique insights, but past predictions for Altjira were generally not well constrained.

It is thus exciting to note that the components of Altjira have just begun their mutual event season! Mutual events can provide detailed information on absolute sizes, shapes, and orientations of the individual objects which then directly constrain albedos and densities. More subtle effects like albedo variegations and phase curves can also be inferred in theory. 

However, we provide a cautionary note that mutual events of faint long-period non-spherical TNOs like Altjira can be very complicated to observe and even harder to interpret. The very similar Manwe-Thorondor binary is a valuable example. Mutual event observations were attempted leading to a complex model with many remaining degeneracies and uncertainties \citep{Rabinowitz2020complex}. Multiple nights of highly precise data (requiring $\gtrsim$4-meter telescopes) at multiple epochs over several years combined with advanced modeling techniques are likely necessary to get detailed information about Altjira. A more modest goal of measuring approximate sizes and shapes may be more realistic. 

A shape model is required for specific details of mutual event predictions like estimated durations, depths, and lightcurve shapes. This is beyond the scope of this paper. However, we can provide the key orbital information needed for future modeling and observing of the Altjira mutual events. 

To begin, we provide an ephemeris for the Altjira system from 2023-2033 in Table \ref{tab:ephem}. This gives our predicted separations and relative positions for the Altjira binary components using our non-Keplerian quadrupole--point-mass model. Uncertainties are calculated using 100 random posterior draws. (We note that this is similar to the methodology used to produce the ``Information Gain'' calculation discussed in Paper II.) This can be used to make predictions for close approaches, although we note that it collapses the uncertainty in time by expressing the uncertainty only in terms of positions. (More detailed model predictions are available from the authors upon request.)

\begin{deluxetable*}{ccCCCCCC}
\tablecaption{System Ephemeris}
\tablehead{
Julian Date & Date & \Delta \alpha \cos{\delta} & \Delta \delta & \sigma_{\Delta \alpha \cos{\delta}} & \sigma_{\Delta \delta} & r & \sigma_{r} \\ 
& & \textrm{('')} & \textrm{('')} & \textrm{('')} & \textrm{('')} & \textrm{('')} & \textrm{('')}}
\startdata
2459945.500 & 2023-01-01 00:00:00 & -0.31319 & 0.10624 & 0.00185 & 0.00141 & 0.33073 & 0.00204 \\
2459945.833 & 2023-01-01 08:00:00 & -0.31180 & 0.10593 & 0.00180 & 0.00139 & 0.32931 & 0.00199 \\
2459946.167 & 2023-01-01 16:00:00 & -0.31036 & 0.10559 & 0.00175 & 0.00136 & 0.32783 & 0.00194 \\
2459946.500 & 2023-01-02 00:00:00 & -0.30885 & 0.10524 & 0.00171 & 0.00134 & 0.32629 & 0.00189 \\
2459946.833 & 2023-01-02 08:00:00 & -0.30729 & 0.10487 & 0.00166 & 0.00132 & 0.32470 & 0.00183 \\
2459947.167 & 2023-01-02 16:00:00 & -0.30568 & 0.10447 & 0.00162 & 0.00130 & 0.32304 & 0.00178 \\
2459947.500 & 2023-01-03 00:00:00 & -0.30400 & 0.10406 & 0.00157 & 0.00127 & 0.32132 & 0.00173 \\
2459947.833 & 2023-01-03 08:00:00 & -0.30227 & 0.10363 & 0.00153 & 0.00125 & 0.31954 & 0.00168 \\
2459948.167 & 2023-01-03 16:00:00 & -0.30047 & 0.10318 & 0.00148 & 0.00123 & 0.31770 & 0.00163 \\
2459948.500 & 2023-01-04 00:00:00 & -0.29862 & 0.10271 & 0.00144 & 0.00121 & 0.31579 & 0.00159 \\ 
\hline
\enddata
\tablecomments{The predicted right ascension and declination positions of Altjira's secondary from 2023 through 2033. Predicted positions, separations, and uncertainties are taken from a sample of 100 random posterior draws of the binary quadrupole--point-mass model. We display the first 10 rows of the table with the rest of the table available as a machine-readable table.}
\label{tab:ephem}
\end{deluxetable*}

Using these values, we can determine the properties of close approaches by interpolation. These are presented in Table \ref{tab:closeapproaches}. The $x$ (e.g., $\Delta RA \cos \delta$) and $y$ positions and velocities and their uncertainties for each close approach are listed. We also list whether the events are superior (primary in front of secondary) or inferior (secondary in front of primary). We note that a close approach of only 4.3 $\pm$ 1.3 milliarcseconds (mas) occurred on Nov 3, 2023, only a short time after our October 23, 2023 HST observation. Using spheres of the nominal sizes from \citet{Vilenius2014tnos}, mutual events are expected when close approaches are less than 7 mas. Since an actual mutual event was likely, we propose that the mutual event season for Altjira has just started.

\begin{deluxetable*}{ccccccccccc}
\tablecaption{Close Approaches for 148780 Altjira}
\tablewidth{0pt}
\tablehead{
Julian Date & Date & Sep. & Sep. err. & $x$ & $x$ err. & $y$ & $y$ err. & $x$ vel. & $y$ vel. & Event Type\\
 & & (mas) & (mas) & (mas) & (mas) & (mas) & (mas) & (mas/hr) & (mas/hr) & }
\startdata
2459972.49306 & 2023-01-27 23:50:00 & 8.327 & 1.262 & 2.714 & 0.851 & 7.872 & 0.932 & 0.743 & -0.256 & Sup. \\
2460112.43403 & 2023-06-16 22:25:00 & 6.287 & 1.253 & 2.103 & 0.850 & 5.925 & 0.921 & 0.713 & -0.253 & Sup. \\
2460165.81944 & 2023-08-09 07:40:00 & 8.602 & 6.980 & -2.823 & 6.192 & -8.126 & 3.223 & -0.386 & 0.135 & Inf. \\
2460252.23611 & 2023-11-03 17:40:00 & 4.272 & 1.339 & 1.449 & 0.912 & 4.019 & 0.981 & 0.747 & -0.266 & Sup. \\
2460304.80208 & 2023-12-26 07:15:00 & 10.374 & 7.322 & -3.328 & 6.490 & -9.826 & 3.390 & -0.409 & 0.139 & Inf. \\
2460391.64236 & 2024-03-22 03:25:00 & 6.708 & 1.334 & 2.298 & 0.904 & 6.302 & 0.980 & 0.724 & -0.261 & Sup. \\
2460444.63542 & 2024-05-14 03:15:00 & 10.381 & 7.221 & -3.475 & 6.390 & -9.782 & 3.364 & -0.381 & 0.135 & Inf. \\
2460531.75347 & 2024-08-09 06:05:00 & 2.781 & 1.370 & 0.956 & 0.941 & 2.612 & 0.996 & 0.721 & -0.265 & Sup. \\
2460585.47569 & 2024-10-01 23:25:00 & 3.871 & 7.713 & -1.307 & 6.805 & -3.643 & 3.632 & -0.393 & 0.143 & Inf. \\
2460671.22222 & 2024-12-26 17:20:00 & 3.699 & 1.443 & 1.295 & 0.987 & 3.465 & 1.053 & 0.746 & -0.274 & Sup. \\
2460723.95486 & 2025-02-17 10:55:00 & 8.925 & 7.773 & -2.997 & 6.855 & -8.406 & 3.663 & -0.397 & 0.142 & Inf. \\
2460810.94097 & 2025-05-15 10:35:00 & 3.774 & 1.419 & 1.327 & 0.973 & 3.532 & 1.032 & 0.710 & -0.267 & Sup. \\
2460864.88889 & 2025-07-08 09:20:00 & 3.666 & 7.869 & -1.302 & 6.919 & -3.427 & 3.748 & -0.375 & 0.142 & Inf. \\
2460950.93750 & 2025-10-02 10:30:00 & 0.193 & 1.505 & 0.077 & 1.045 & 0.177 & 1.083 & 0.735 & -0.278 & Sup. \\
2461004.59375 & 2025-11-25 02:15:00 & 1.669 & 8.377 & -0.573 & 7.353 & -1.568 & 4.014 & -0.400 & 0.149 & Inf. \\
2461090.26736 & 2026-02-18 18:25:00 & 2.939 & 1.527 & 1.069 & 1.051 & 2.738 & 1.107 & 0.731 & -0.278 & Sup. \\
2461143.67014 & 2026-04-13 04:05:00 & 5.150 & 8.227 & -1.832 & 7.214 & -4.813 & 3.955 & -0.380 & 0.145 & Inf. \\
2461230.29167 & 2026-07-08 19:00:00 & 0.170 & 1.533 & 0.052 & 1.069 & 0.162 & 1.098 & 0.710 & -0.276 & Sup. \\
2461284.88542 & 2026-09-01 09:15:00 & 2.570 & 8.624 & 0.944 & 7.537 & 2.390 & 4.191 & -0.378 & 0.150 & Inf. \\
2461369.96875 & 2026-11-25 11:15:00 & 1.022 & 1.634 & -0.384 & 1.142 & -0.948 & 1.168 & 0.743 & -0.289 & Sup. \\
2461423.55556 & 2027-01-18 01:20:00 & 0.817 & 8.894 & -0.294 & 7.767 & -0.762 & 4.333 & -0.396 & 0.153 & Inf. \\
2461509.46875 & 2027-04-13 23:15:00 & 0.939 & 1.609 & 0.320 & 1.121 & 0.883 & 1.154 & 0.713 & -0.282 & Sup. \\
2461563.84722 & 2027-06-07 08:20:00 & 0.961 & 8.798 & 0.354 & 7.666 & 0.893 & 4.318 & -0.368 & 0.149 & Inf. \\
2461649.56944 & 2027-09-01 01:40:00 & 3.150 & 1.675 & -1.165 & 1.186 & -2.927 & 1.182 & 0.720 & -0.289 & Sup. \\
2461704.32639 & 2027-10-25 19:50:00 & 6.594 & 9.385 & 2.484 & 8.157 & 6.108 & 4.642 & -0.387 & 0.157 & Inf. \\
2461788.95833 & 2028-01-18 11:00:00 & 1.462 & 1.738 & -0.532 & 1.222 & -1.361 & 1.236 & 0.737 & -0.296 & Sup. \\
2461842.89583 & 2028-03-12 09:30:00 & 1.230 & 9.320 & 0.448 & 8.094 & 1.146 & 4.621 & -0.381 & 0.155 & Inf. \\
2461928.79861 & 2028-06-06 07:10:00 & 2.255 & 1.714 & -0.845 & 1.214 & -2.090 & 1.210 & 0.703 & -0.289 & Sup. \\
2461984.05903 & 2028-07-31 13:25:00 & 7.929 & 9.520 & 3.123 & 8.242 & 7.288 & 4.764 & -0.366 & 0.156 & Inf. \\
2462068.69097 & 2028-10-24 04:35:00 & 5.327 & 1.825 & -2.060 & 1.305 & -4.913 & 1.275 & 0.733 & -0.302 & Sup. \\
2462123.31944 & 2028-12-17 19:40:00 & 8.083 & 10.018 & 3.108 & 8.660 & 7.461 & 5.035 & -0.390 & 0.162 & Inf. \\
2462208.05208 & 2029-03-12 13:15:00 & 2.484 & 1.823 & -0.954 & 1.293 & -2.294 & 1.285 & 0.719 & -0.299 & Sup. \\
2462262.81597 & 2029-05-06 07:35:00 & 5.915 & 9.801 & 2.330 & 8.459 & 5.436 & 4.950 & -0.365 & 0.157 & Inf. \\
2462348.13194 & 2029-07-30 15:10:00 & 5.862 & 1.851 & -2.283 & 1.333 & -5.399 & 1.284 & 0.707 & -0.299 & Sup. \\
2462487.69097 & 2029-12-17 04:35:00 & 6.153 & 1.956 & -2.404 & 1.406 & -5.664 & 1.359 & 0.737 & -0.312 & Sup. \\
2462542.36111 & 2030-02-09 20:40:00 & 8.972 & 10.482 & 3.569 & 9.011 & 8.231 & 5.353 & -0.381 & 0.165 & Inf. \\
2462627.30556 & 2030-05-05 19:20:00 & 4.850 & 1.916 & -1.929 & 1.377 & -4.450 & 1.332 & 0.703 & -0.303 & Sup. \\
2462767.35764 & 2030-09-22 20:35:00 & 8.907 & 2.012 & -3.532 & 1.468 & -8.177 & 1.376 & 0.719 & -0.313 & Sup. \\
2462906.70139 & 2031-02-09 04:50:00 & 6.577 & 2.055 & -2.633 & 1.485 & -6.027 & 1.419 & 0.725 & -0.316 & Sup. \\
2463046.64583 & 2031-06-29 03:30:00 & 8.236 & 2.039 & -3.359 & 1.490 & -7.519 & 1.392 & 0.697 & -0.310 & Sup. \\
2463186.42708 & 2031-11-15 22:15:00 & 10.624 & 2.169 & -4.292 & 1.594 & -9.718 & 1.471 & 0.730 & -0.325 & Sup. \\
2463325.85069 & 2032-04-03 08:25:00 & 7.932 & 2.140 & -3.276 & 1.562 & -7.224 & 1.463 & 0.706 & -0.318 & Sup. \\
2463465.95139 & 2032-08-21 10:50:00 & 11.765 & 2.197 & -4.866 & 1.628 & -10.712 & 1.474 & 0.704 & -0.322 & Sup. \\
2463605.40972 & 2033-01-07 21:50:00 & 11.135 & 2.293 & -4.626 & 1.691 & -10.129 & 1.549 & 0.728 & -0.333 & Sup. \\
2463745.14583 & 2033-05-27 15:30:00 & 10.635 & 2.245 & -4.486 & 1.661 & -9.643 & 1.510 & 0.693 & -0.322 & Sup. \\
2463885.11806 & 2033-10-14 14:50:00 & 14.455 & 2.370 & -6.134 & 1.773 & -13.089 & 1.573 & 0.717 & -0.335 & Sup. \\ 
\enddata
\tablecomments{Close approaches are defined as events where the primary and secondary are separated by less than 15 mas, not necessarily implying a mutual event. $x$ and $y$ correspond to $\Delta \alpha \cos{\delta}$ and $\Delta \delta$, respectively. Note that timing and depth uncertainty are significant for inferior events. \label{tab:closeapproaches} }
\end{deluxetable*}

We calculated the close approach predictions using our Keplerian model and our non-Keplerian model to demonstrate that non-Keplerian effects are necessary for accurate mutual event predictions. For example, the separations predicted for the same event differed by $\sim$3 mas for superior events and $\sim$5 mas for inferior events, comparable to the sizes of the components. The mutual event season is about 1.5 years earlier in the non-Keplerian model and centered roughly on 2026, which partially explains why our predicted mutual events have started so soon. Timing predictions differed by about 0.5 hour and growing to 1.5 hours for superior events, comparable to the estimated timing uncertainties. However, for inferior events, the timing predictions were different between Keplerian and non-Keplerian by 32-50 hours! Inferior events happen near apoapse, so the precession included in non-Keplerian modeling has a larger lever arm; furthermore, the velocity is slower so that a larger distance in position corresponded to an even larger difference in timing. We conclude that including non-Keplerian effects can make a significant difference in mutual event predictions. 

Given the sensitivity of the predictions to small effects, it is also worth mentioning that unmodeled systematic uncertainties could lead to significant offsets from our predicted close approaches. Possible systematic uncertainties are not included in Table \ref{tab:closeapproaches}. 

Shadowing events are not shown in Table \ref{tab:closeapproaches}. Calculation of the shadowing events (based on predictions for close approaches as seen from the perspective of the Sun) shows that they are about $\pm$1 mas separated from the occultations and occur at a similar time for superior events and about two hours earlier for inferior events. Differences in Keplerian and non-Keplerian predictions for shadowing are very similar to the differences seen in occultations. 

We can provide some context about the expected depth and duration of mutual events assuming different shape models. First, we note that since the two components have similar brightness that a perfect shadowing and/or occultation would lead to a $\sim$50\% drop in flux for an object that is typically $V \simeq 23$ magnitudes in brightness. Non-zero impact parameters and non-spherical shapes suggest that this would be rare; $\sim$10\% drops in flux should be common for the closest approaches. Since the objects are probably similar in size, these drops should generally be gradual, e.g., ingress and egress times are about half of the event durations. The fastest expected brightness changes are $\sim$0.1 mag/hr.

To estimate durations, we begin by noting that Altjira's typical distance is 45 AU, where 1 mas corresponds to about 32 km. Taking the sizes from \citet{Vilenius2014tnos}, this gives radii of approximately 3.8$^{+0.6}_{-2.2}$ and 3.4$^{+0.5}_{-2.0}$ mas, but this assumes low equal albedos, equal densities, and spherical shapes. Perhaps this is representative of the long axis of a non-spherical shape. If we assume spherical shapes and -1-$\sigma$ radii of 1.6 and 1.4 mas then a grazing event requires a close approach distance of only 3.0 mas. About 10 such events are predicted, most of which have uncertainties that are reasonably small so that the statistical expectation is a $\sim$95\% chance of some kind of event (especially when including shadowing). There is significant correlation in the predicted close approach distances so that clear observation and interpretation of an earlier event could shrink uncertainties in later events. New additional astrometric measurements may help somewhat, especially later in the mutual event season. Conversely, precise mutual events with clear interpretations can provide effective astrometric observations. 

Continuing with the small sphere model (1.6 and 1.4 mas radii) and using the velocities given in Table \ref{tab:closeapproaches}, a central superior event would last about 3.5 hours and a central inferior event would last about 7 hours. Non-central events are correspondingly shorter, though multiple events would have $\gtrsim$1 hour expected durations. If the objects are bigger (in some direction), then these durations grow correspondingly. 

All things considered, about $\sim$10 close approaches are likely to yield strong mutual events. Several other close approaches could lead to detectable mutual events if the components are larger than expected, non-spherical, or composed of close binaries as we have hypothesized. 

If one or both of the components are a close binary, this changes mutual event predictions because the components are smaller (e.g., 1.1 mas if equal-mass, equal-albedo, equal-density spheres) and because this inner binary would have orbital motion that could be a few times faster than the outer binary motion. This is similar to how an elongated object rotating on a similar timescale to the mutual event duration can significantly affect the shape of the resulting lightcurve. However, the inner binary separation would also allow for mutual events at larger separations. 

Furthermore, because the inner binary is so close together, mutual events between inner binary components would be much more likely. If the inner and outer binaries are even reasonably closely aligned, similar to what is seen in our fits, inner mutual events could also be happening now. Depending on the separation and orientation of the inner binary components, light curve observations could potentially distinguish between a close binary, a contact binary, and a single triaxial ellipsoid. 

We suggest that the next step in preparing for Altjira's mutual events would be a detailed lightcurve of the system. Such a lightcurve could detect inner mutual events or at least provide information on the amplitudes, periods, and shapes of the two components. This information is a prerequisite to precise mutual event model prediction and interpretation. We note that the binary components are sometimes separated by 0.3 arc seconds which could lead to resolved lightcurves with high-quality data under great seeing with PSF modeling. The ephemeris in Table \ref{tab:ephem} can be used to predict such ideal observational times.

As a final word of caution, we note that our ephemeris and predictions depend on our model assumptions. It is possible that a similar difference between the Keplerian and non-Keplerian models would also be found in different non-Keplerian models (such as a two-quadrupole model). Given the statistical and systematic uncertainties, observing for as long as possible before and after the nominal events is prudent. 

We also speculate that inaccurate orbital modeling may be a major contributing factor to the fact that mutual events have not yet fulfilled their promise of improved characterization of other TNO binaries \citep[e.g.,]{2009AJ....137.4766R,Rabinowitz2020complex}. For example, the error in the ephemeris for Haumea identified in \citet{proudfoot2024haumea} (Paper III) could have contributed to the challenges of interpreting observed mutual events, though Hi'iaka's unexpectedly strong lightcurve \citep{Hastings2016unexpected} also contributed. 

\section{Discussion} \label{sec:discussion}

Given the observed non-Keplerian motion in Altjira, what can we say about its shape? As noted above, $J_2R^2 = 17000^{+9000}_{-7000}$ km$^2$ is not precisely detected and is unrealistically assumed to be entirely attributable to the primary (with a point-mass secondary). Still, to guide future interpretation, we can comment on the implications of these measurements for the shape of Altjira's primary. 

To convert a measurement of $J_2R^2$ to a particular three-dimensional mass configuration requires assuming a shape model. For example, an oblate ellipsoid model for Altjira that matches the nominal $J_2R^2$ would require an equatorial radius $\sim$2.5 times larger than the polar radius. This disk-like shape is unrealistic (e.g., never seen in real objects), suggesting a different shape model would be more appropriate. 

We consider a Cassinoid as a flexible model that approximates a contact binary while being similar to observed small bodies. \bp{A Cassinoid is a dumbbell-like shape, closely representing the dumbbell figures of a rotating, self-gravitating fluid body. It is very useful as it has a simple algebraic representation that can easily be used to calculate moments of inertia analytically. See the Appendix of \citet{descamps2015dumb} for more information about Cassinoids.} (We note that a Roche model that assumes fluid strength-less interiors is not an ideal physical characterization of small TNO components.) For the nominal $J_2R^2$ and mass of Altjira, we find that the minimum size of the Cassinoid would require a density of 0.2 g cm$^{-3}$, a very low value and also somewhat inconsistent with the thermal size measurements. Considering uncertainties suggests that a Cassinoid model has a $\sim$20\% chance of being an acceptable description of the observations. 

To reach the higher $J_2R^2$ values implied by our fits requires rejecting the contact binary model and adopting components that are physically separated. For example, two spheres with density of 0.5 g cm$^{-3}$ (1 g cm$^{-3}$) could be explained by components of $\sim$85 km ($\sim$70 km) in radius, separated by 300-450 km. This would also be more consistent with the thermal models, assuming that these measure total surface area. We note again that a close hierarchical triple configuration was found to be an excellent fit to the observations. 

The results of \citet{correia2018chaotic} and Paper I for the Lempo hierarchical triple system show that the current model is dynamically unstable on $\sim$10$^5$-year timescales. Examining the stability of an Altjira hierarchical triple is beyond the scope of this work, though we can confirm that it appears very stable on the  timescale of the observations. Altjira is also much more hierarchical than Lempo ($a_{\textrm{out}}/a_{\textrm{in}} \simeq 40$ instead of 5.5 for Lempo) which should significantly increase its long-term stability.

More advanced modeling of the shape of the components of Altjira would be justified once lightcurve modeling (and then mutual events) can provide additional constraints on relative sizes and shapes. For example, for any reasonable set of parameters, the proposed tight inner binary should be tidally locked (assuming no major perturbations from the outer component), so that the observed lightcurve period would be equivalent to period of the inner binary. 


This nominal configuration is relatively similar to Lempo, although the ratio of the outer and inner binary separations is larger. Without knowledge of how the mass is distributed between the inner and outer binary -- which is not well-determined in our model -- we cannot directly assess how the angular momentum is distributed among the components. However, it appears that, unlike the Lempo system, most of the angular momentum for Altjira is in the outer binary. Whether such a configuration is a reasonable outcome of gravitational collapse triggered by SI requires more advanced collapse models.  

Another formation model to consider for cold classical hierarchical triples is formation by binary-binary encounters \citep[e.g.][]{2020MNRAS.499.4206B}. Such encounters are relatively common for binaries as wide as Altjira and are not adequately studied. Like Lempo, Altjira's inner and outer binaries are relatively well-aligned which is not a common outcome of these encounters, though additional modeling is also needed to better understand this scenario. 

\section{Conclusions} \label{sec:conclusions}

Altjira is proposed to be cold classical TNO binary with an unusually high inclination. Such cold classical binaries are proposed to be formed through gravitational collapse after the streaming instability. Altjira's configuration is consistent with these models and provides motivation for improved models that can resolve the angular momentum contribution of individual TNO binary components. It is also possible that Altjira formed or was modified by binary-binary encounters. 

We have modeled the non-Keplerian orbit of TNO binary (148740) Altjira (2001 UQ$_{18}$), using the newest HST data from 2023. We analyzed the images with our precise Point Spread Function (PSF) fitting routine called \texttt{nPSF}, as presented in this paper. Using the astrometry output from \texttt{nPSF} and astrometry from past HST visits and Keck data between 2006 and 2020, we modeled Altjira in different configurations with our Bayesian parameter inference tool \texttt{Multimoon} (Paper I). 

We confirmed that a non-Keplerian model for Altjira was preferred over a Keplerian model at the $\sim$2.5-$\sigma$ level. A binary quadrupole--point-mass model finds an oblateness of $J_2R^2 = 17000^{+9000}_{-7000}$ km$^2$ which is larger than expected for the size of Altjira. Even considering that the non-Keplerian effects are likely due to both shapes, a triaxial ellipsiod and Cassinoid (dumbbell) model are not plausible for the majority of our uncertainty region. We thus propose that one or both of the components of Altjira are near-equal mass unresolved ``inner'' binaries. Our best fit to the data is a hierarchical triple configuration, though we did not explore this parameter space in detail. We thus conclude that Altjira is likely a unresolved hierarchical triple. 

We also call attention to Altjira's ongoing mutual event season. After obtaining detailed lightcurves (ideally resolved), mutual event observations could provide significantly improved understanding of the shapes, sizes, albedos, and densities of the components of this system. We provide an ephemeris for close approaches in this system -- which is different from the prediction using Keplerian orbits -- but leave detailed predictions for future work. We note that mutual event interpretations are challenging as seen for the very similar Manwe-Thorondor system \citep{Rabinowitz2020complex}. 

A self-consistent analysis of physical properties of the components of Altjira that combines mass and $J_2R^2$ measurements, thermal contraints, lightcurve shapes, and eventually mutual event data would be able to determine detailed information about the sizes, shapes, and configurations of this primordial likely-triple TNO.

\begin{acknowledgments}
\emph{Acknowledgments.} \qquad We acknowledge the support of Dallin Spencer with various aspects of \texttt{MultiMoon} fitting and general help. We acknowledge many cohorts of Brigham Young University Physics 529 students who worked on or tested various versions of \texttt{nPSF} including Kyle Adams, Crystal-Lynn Bartier, Scott Call, Ian Clark, Jared Davidson, Jarrod Hansen, Jacob Jensen, Daniel Jones, Emma Rasmussen, Rochelle Steele, Savanah Turner, Nicholas Wallace, and Denzil Watts. 

This research is based on observations made with the NASA/ESA Hubble Space Telescope obtained from the Space Telescope Science Institute, which is operated by the Association of Universities for Research in Astronomy, Inc., under NASA contract NAS 5–26555. These observations are associated with program 17206. Support for this work was funded by grants associated with this program. \textbf{Some of the data presented in this article were obtained from the Mikulski Archive for Space Telescopes (MAST) at the Space Telescope Science Institute. The specific observations analyzed can be accessed via \dataset[DOI: 10.17909/ht9x-nt22]{https://doi.org/10.17909/ht9x-nt22}}

Some of the data presented herein were obtained at Keck Observatory, which is a private 501(c)3 non-profit organization operated as a scientific partnership among the California Institute of Technology, the University of California, and the National Aeronautics and Space Administration. The Observatory was made possible by the generous financial support of the W. M. Keck Foundation. 

The authors wish to recognize and acknowledge the very significant cultural role and reverence that the summit of Maunakea has always had within the Native Hawaiian community. We are most fortunate to have the opportunity to conduct observations from this mountain. 

\end{acknowledgments}

\bibliography{sample631}{}
\bibliographystyle{aasjournal}


\end{document}